\documentstyle[pre,aps,epsf,psfig]{revtex}
\newcommand \be[1]{\begin{equation}\label{#1}}
\newcommand \ee{\end{equation}}
\newcommand \beq[1]{\begin{eqnarray}\label{#1}}
\newcommand \eeq{\end{eqnarray}}
\newcommand \Beq{\begin{eqnarray}}
\newcommand \Eeq{\end{eqnarray}}
\newcommand \bib{\bibitem}

\newcommand{\DK}[1]{\mbox{\boldmath$#1$}}

\newcommand \al{\alpha}
\newcommand \nn{\nonumber}
\newcommand \om{\omega}
\newcommand \eps{\epsilon}
\newcommand \veps{\varepsilon}
\newcommand \arsinh{{\rm arsinh}}
\newcommand \sh{{\rm sinh}}
\newcommand \ch{{\rm cosh}}
\newcommand \th{{\rm tanh}}
\begin{document}

\draft
\title{
Relativistic semiclassical approach in strong-field nonlinear photoionization 
}

\author{J. Ortner $\,^{a)}$ and V.~M.~Rylyuk $\,^{b)}$}
\address{
$\,^{a)}$ {\it Institut f\"ur Physik,Humboldt Universit\"at zu Berlin, 
Invalidenstr. 110, 10115 Berlin, Germany}\\
$^{b)}${\it Department of Theoretical Physics, University of Odessa, 
Dvorjanskaja 2, 270100, Odessa, Ukraine}}
\date{to be published in Phys. Rev. A}
\maketitle
\begin{abstract}
Nonlinear relativistic ionization phenomena induced by a strong laser radiation with elliptically polarization are considered. The starting point is the classical relativistic action for a free electron moving in the electromagnetic field created by a strong laser beam. The application of the relativistic action to the classical barrier-suppression ionization is briefly discussed. Further the relativistic version of the Landau-Dykhne formula is employed to consider the semiclassical sub-barrier ionization. Simple analytical expressions have been found for: (i) the rates of the strong-field nonlinear
ionization including relativistic initial and final state effects; (ii) the most probable value of the components of the photoelectron final state momentum;  (iii) the most probable direction of photoelectron emission and (iv) the distribution of the photoelectron momentum near its maximum value.
\end{abstract}


\section{Introduction}

Relativistic ionization phenomena induced by strong laser light have become a topic of current interest \cite{R90,CR94,PMK97,PMK98,DK98,K98,CR98,K99}. In the nonrelativistic theory it is assumed that the electron velocity in the initial bound state as well as in the final state is small compared with the speed of light. However, the electrons may be accelerated up to relativistic velocities in an intense electromagnetic field produced by modern laser devices. If the ponderomotive energy of the electron is of the order of the rest energy a relativistic consideration is required. Relativistic effects in the final states become important for an infrared laser at intensities of some $10^{16} {\rm W}\,{\rm cm}^{-2}$.  The minimal intensity required for relativistic effects increases by two orders of magnitude for wavelength corresponding to visible light. Ionization phenomena connected with relativistic final state effects have been studied for the cases of linearly and circularly polarized laser radiation both in the tunnel \cite{K98,K99} and above-barrier regimes \cite{R90,CR94,CR98}. The main relativistic effects in the final state are \cite{R90,CR94,DK98,K98,CR98,K99}: (i) the relativistic energy distribution and (ii) the shift of the angular distribution of the emitted electrons towards the direction of propagation of incident laser beam. It has been shown that a circularly polarized laser light produces a large amount of relativistic electrons \cite{R90,CR94,K99}. On the contrary, it has been found that the ionization rate for relativistic electrons is very small in the case of linear polarization \cite{K98}.

Relativistic effects have also to be taken into account if the binding energy $E_b$ in the initial state is comparable with the electron rest energy \cite{PMK97,PMK98}. A relativistic formulation is necessary for the ionization  of heavy atoms or singly or multiply charged ions from the inner K-shell. In Refs. \cite{PMK97,PMK98} the relativistic version of the method of imaginary time has been employed to calculate the ionization rate for a bound system in the presence of intense static electric and magnetic fields of various configurations. Analytical expressions have been found which apply to nonrelativistic bound systems as well as to initial states with an energy corresponding to the upper boundary of the lower continuum.
   
The present paper is aimed to consider the nonlinear photoionization connected with  relativistic final states velocities and/or low lying initial states  from a unique point of view. 
Further the current work extends the investigation of relativistic ionization phenomena to the case of arbitrary {\em elliptical} polarization. 

Certainly this may be done in the framework of the so-called strong field approximation \cite{R80}. In the papers of Reiss and of Crawford and
Reiss  \cite{R90,CR94,CR98} a relativistic version of this  approximation has been given for the ionization of an hydrogen atom with linearly and circularly polarized light.
 Within this approximation one calculates the transition amplitude between the initial Dirac state for the hydrogen atom and the final state described by the relativistic Volkov wave function. Coulomb corrections are neglected in the Volkov state.  Therefore the final results are obtained only within exponential accuracy. Analytical
results for the ionization rate applying to above barrier cases as well as to tunneling cases have been given in Refs. \cite{R90,CR94,CR98}.  However, the corresponding expressions are complicated and contains infinite sums over all multiphoton processes. Numerical calculations are needed to present the final results. 

In contrast to the more sophisticated investigations, such as the solution of the Dirac equation or the strong field approximation, we are aimed to obtain simple analytical expressions. From our final formulas  the explicit dependence of the ionization rate and of the photoelectron spectrum on the parameters, such as binding energy of the atom, field strength, frequency and ellipticity of the laser radiation may be understood without the need of numerical calculations. In this sense our approach resembles that of Popov {\em et al.} \cite{PMK97,PMK98} and of Krainov \cite{K98,K99}.

\section{Relativistic action and classical barrier-supression ionization}\label{BSI}

Let us start with the classical relativistic action for an electron of charge $e$ moving in the field of an  electromagnetic plane wave with the vector potential $\DK{A}(t-x/c)$. Here and below $\DK{A}$ denotes a two-dimensional vector in the y-z plane. The action may be found as a solution of the Hamilton-Jacobi equation and reads \cite{Landau}
\be{Srel}
S_f(\xi;\xi_0)= mc^2\Biggl\{ \DK{f} \cdot \frac{\DK{r} }{c}-\alpha \frac{x}{c}-\frac{1+\al^2+f^2}{2 \al} \left(\xi-\xi_0\right)+ \frac{e}{m c^2 \al} \DK{f} \int_{\xi_0}^{\xi} \DK{A} d\xi - \frac{e^2}{2 m^2c^4 \al} \int_{\xi_0}^{\xi} \DK{A}^2 d\xi \Biggr\} \,,
\ee
where $\al$ and $\DK{f}=(a_1,a_2)$ are constants, $\DK{r}=(y,z)$; further is $\xi=t-{x}/{c}$, $\xi_0$ is the initial value. Assuming a harmonic plane wave of elliptic polarization with the electric field $\DK{E}=F\left[{\DK{e}_y\cos{\om\xi}+ g\DK{e}_z\sin{\om\xi}}\right]$ we find the following expression for the relativistic action
\beq{Srel2}
S_f(\xi;\xi_0)&=& mc^2\Biggl\{ \DK{f} \cdot \frac{\DK{r} }{c}-\frac{\alpha}{2}\left(t+ \frac{x}{c}\right)-\frac{\beta^2}{2 \al} \left(\xi-\xi_0\right)+\frac{\eps}{\alpha \om} \left[ a_1\left( \cos \om \xi - \cos \om \xi_0\right) \right. \nn \\
&& \left. + a_2 g \left( \sin \om \xi - \sin \om \xi_0 \right) \right] - \frac{\eps^2}{8 \alpha \om} \left(g^2-1\right)\left( \sin 2 \om \xi - \sin 2 \om \xi_0 \right) \Biggr\} \,,
\eeq
where the notation $\beta^2 =1+a_1^2+a_2^2 +((1+g^2)/2)\eps^2$ has been introduced,  the parameter $\eps=e F/\om m c $ characterizes the strength of relativistic effects. Further the vector potential of the laser radiation has been choosen in the form 
\be{pot}
A_x=0\,,~~~A_y=-\frac{cF}{\om} \sin \om \xi\,,~~~A_z=g \frac{cF}{\om} \cos \om \xi\,.
\ee

By applying the usual Hamilton-Jacobi method we take the derivative of the action  $S_f$ with respect to the constants  $a_1,a_2$ and $\alpha$ and set the result equal to new constants $\beta_1,\beta_2$ and $\beta_3$ in order to obtain the electron trajectory under the influence of the wave field.  We obtain that the electron motion in the field and in the laboratory coordinate system is given by the equations ($\xi_0=0$),
\beq{trajec}
\alpha^2 (t+x/c) - \beta^2 \xi+\frac{2 \eps}{ \om}
\left(a_1\cos{\om\xi}+ ga_2\sin{\om\xi}\right)
+\frac{1-g^2}{4 \om} \eps^2 \sin{2\om\xi}= \beta_3&,&~~~
v_x = c \frac{f(\xi) -1}{f(\xi) +1} \,,\nn\\
y = \beta_1 +\frac{ca_1}{\alpha} \xi -\frac{c \eps}{ \alpha\om}
\cos{\om\xi}\,&,&~~~
v_y =\frac{2c}{ \alpha (1+f(\xi))} \left\{a_1 +
\eps  \sin \om \xi \right\},\nn \\
z =\beta_2 +\frac{ca_2}{\alpha} \xi -g\frac{c \eps}{ \alpha\om}
\sin{\om\xi}\,&,&~~~
v_z =\frac{2c}{ \alpha (1+f(\xi))} \left\{a_2 -
\eps g \cos \om \xi \right\}\,,\nn\\
f(\xi)= \frac{\delta ^2}{\alpha^2} +
\frac{2\eps}{\alpha^2} \left( a_1\sin{\om\xi}-
g a_2\cos{\om\xi}\right)+ \frac{1-g^2}{\alpha^2} \eps^2
\sin^2{\om\xi}&,&
\eeq
where $\beta_1,\beta_2$ and $\beta_3$ together with $a_1,a_2$ and $\alpha$ have to be determined from the initial conditions for position and velocity. Further we have introduced the notation $ \delta ^2 =1+a^2_1 + a^2_2 + g^2 \eps ^2$. 

Quantum effects may be neglected, for strong enough fields, i.e., $F \gg F_B= E_0^2/4Z$ (in a.u., where $E_0$ is the electron energy in the initial state and $Z$ is the effective charge of the atomic core). In this case the ionization process may be described by an electron trajectory given in Eqs. (\ref{trajec}). In a pure classical task the constants of motion may be determined from the initial velocity and  position of the electron at the beginning of the laser action. However, the initial state is given by quantum mechanics. According to a simple classical picture of ionization, the barrier supression ionization (BSI) \cite{CBB92}, the transition occurs from the bound state to that continuum state which has zero velocity at the time $t$ with the phase $\xi$ of the vector potential $\DK{A}(\xi)$.  From this condition we have to choose the constants as
\beq{semi}
\alpha&=&\sqrt{\delta ^2 + 2 \eps \left(a_1 \sin \om \xi -g a_2 \cos \om \xi \right) + (1-g^2) \eps^2 c^2 \sin^2 \om \xi} \,,\nn \\  
a_1&=&- \eps \sin \om \xi\,,~~~
a_2=g \eps \cos \om \xi\,.
\eeq
The maximal ionization rate occurs at the maximum of the electric field of the laser radiation. 
For our choice of the gauge (see Eqs. (\ref{pot})) the electric field has its maximum at the phase $\xi=0$. (For the sake of simplicity we neglect the second maximum at $\xi=\pi$). From Eqs. (\ref{semi}) we conclude that the most probable final state is described by the constants
\be{semip}
\alpha=1\,,~~~a_1=0\,,~~~a_2=\eps g\,,
\ee
 Two important results follow from this derivation. 

First, consider the components of the final electron drift momentum along the beam propagation, $p_x=c\,(1-\alpha^2+a_1^2+a_2^2)/2 \alpha$, along the major axis, $p_y=c\, a_1$, and along the small axis of the polarization ellipse, $p_z=c \, a_2$, respectively. From Eq. (\ref{semip}) we see that the photoelectrons are preferably produced with the drift momentum
\be{semipp}
p_x= \frac{\eps^2 g^2}{2} c\,,~~~p_y=0\,,~~~p_z=\eps g c\,.
\ee

 For a laser wavelength of 
$\lambda  = 780$ nm and for laser intensities of about
${\rm I} = (10^{18} - 10^{19}) ~{\rm W}/{\rm cm}^2$, the parameter $\eps $ is equal to:
$\eps_1 = 0.65 -2.1$ for the linearly polarized wave ($ g=0 $) and
$\eps_2 = 0.46 -1.46$ for the circularly polarized wave ($ g^2=1 $).
According to the classical barrier-supression ionization model the photoelectrons are
emitted  with a relativistic drift momentum \cite{CBB92} at these laser intensities and for sufficiently large ellipticity $g$ . On the contrary, in the case of linear polarization the photoelectrons have a zero drift momentum.

Second, the angle between the electron drift momentum components along and perpendicular to the direction of the laser beam propagation is shifted toward the forward direction and reads
\be{angle}
\tan \theta=\frac{p_{\perp}}{|p_z|}=\frac{|p_{x}|}{|p_{z}|}=\frac{\eps |g|}{2} \,,
\ee

For linearly polarized laser light we obtain $\tan \theta = 0$.  For the case of circularly polarized light (where $\tan \theta = \eps/2$) our result coincide with that of previous works \cite{CR94,K99}.

\section{Relativistic semiclassical approach}
Consider now the process of nonlinear ionization of a strongly bound electron with a binding energy $E_b$ comparable with the rest energy. Recently  the ionization process in static crossed electric and magnetic fields has been considered \cite{PMK97,PMK98}. The results of this paper may be applied to the ionization in laser fields only for the case of very strong fields $\eps \gg 1$. With an increasing frequency of the laser light (especially for a tentative x-ray laser) very high laser intensities are required to satisfy this condition. Therefore it is necessary to generalize the result of \cite{PMK97,PMK98} to the case of nonzero frequencies. We consider the sub-barrier ionization. The condition to be satisfied is the opposite to the case of pure classical ionization, $F \ll F_B$, in addition we have the quasiclassical condition $\hbar \om \ll E_b$. No restrictions are applied to the parameter $\eps$. Thus we will cover both the regime of relativistic tunnel and multiphoton ionization. 

We employ the relativistic version of the Landau-Dykhne formula \cite{PMK97,DK98}. The ionization probability in quasiclassical approximation and with exponential accuracy reads
\be{prob}
{W} \propto \exp\left\{- \frac{2}{\hbar} \, {\rm Im}~\left(S_f(0;t_0)+S_i(t_0)\right)\right\} \,,
\ee
where $S_i=E_0 t_0$ is the initial part of the action, $S_f$ is given by Eq. (\ref{Srel}) (or  Eq. (\ref{Srel2})). The complex initial time $t_0$ has to be determined from the classical turning point in the complex half-plane \cite{PMK97,DK98}:
\be{cond1}
E_f(t_0)=mc^2\Biggl\{\frac{1+\al^2+f^2}{2 \al} - \frac{e}{m c^2 \al} \DK{f} \DK{A(t_0)}  + \frac{e^2}{2 m^2c^4 \al}  \DK{A}^2(t_0)  \Biggr\}=E_0=mc^2-E_b \,.
\ee
Explicitely we obtain for $\lambda_0=-i \om t_0$ the following relation in the case of an elliptically polarized planar wave,
\be{lambda0}
\sh^2\lambda_0-g^2\left(\ch\lambda_0 -
\frac{\sh\lambda_0}{\lambda_0}\right)^2= \gamma^2(\alpha),
\ee
where $\gamma(\alpha) =\eta \sqrt{1+ \alpha^2 -2\alpha \varepsilon_0}$,
or $\gamma^2(\alpha) = (1-\alpha)^2 \eta^2 +\alpha \gamma^2$, with the dimensionless initial energy $\veps_o=E_0/mc^2$ and the relativistic adiabatic parameter $\eta = \eps^{-1}= \om mc/eF$.
Eq. (\ref{prob}) together with Eqs. (\ref{Srel2}) and (\ref{lambda0}) expresses the transition rate  between the initial state and the final Volkov state with abitrary momentum within exponential accuracy. It applies for the case of sub-barrier ionization with elliptically polarized laser light.

We are now interested in the {\em total} ionization rate. Within exponential accuracy it suffices to find the maximum of the transition rates between initial state and all possible final states. Aquivalently, one has to find the minimum of the imaginary part of the action as a function of the final state momentum.
The minimization of the imaginary part of the action with respect to the components of the final state momentum leads to the following boundary conditions \cite{PKM67}
\be{cond2}
(x,\DK{r})(t_0)=0\,,~~~
{\rm Im}~(x,\DK{r})(t=0)=0\,.
\ee
From these conditions we obtain that the most probable final state is characterized by the parameters
\Beq
\label{alpha} \alpha^2 &=& 1+\frac{1}{2\eta^2} \left(1+g^2 -2g^2
\frac{\sh^2\lambda_0}{\lambda_0^2} - \frac{1-g^2}{2\lambda_0}
\sh2\lambda_0 \right)\,,\\
\label{a1} a_1&=&0\, \\
\label{a2} a_2&=&(g/\eta)(\sh\lambda_0/\lambda_0)\,.
\Eeq

 Substituting the values $\lambda_0=-i \om t_0$ and $\al$ into the final state action we obtain the probability of relativistic quasiclassical ionization in the field of elliptically polarized laser light. Within exponential accuracy we get
\be{relprob}
W \propto \exp\left(-\frac{2E_b}{\hbar \omega} f(\gamma, g, E_b) \right),
\ee
where
\be{fg}
f(\gamma, g, E_b) = \left(1+\frac{1+g^2}{2 \gamma^2 \alpha}+\frac{mc^2}{E_b}
\frac{(1-\alpha)^2}{2\alpha} \right) \lambda_0 -
\left( 1-g^2 + 2g^2 \frac{\th\lambda_0}{\lambda_0}\right)
\frac{\sh2\lambda_0}{4\gamma^2\alpha}\,.
\ee
The magnitudes $\al$ and $\lambda_0$ has to be taken as the solution of Eqs. (\ref{lambda0}) and (\ref{alpha}). Further $\gamma = \sqrt{2mE_b}\om/eF$ is the common adiabatic Keldysh parameter from nonrelativistic theory \cite{DK98}. Equation (\ref{relprob}) is the most general expression for the relativistic ionization rate in the quasiclassical regime and for field strength smaller than the above-barrier threshold. It describes  both the tunnel as well as the multiphoton ionization. It is the relativistic generalization of previous nonrelativistic results \cite{PKM67}. 

\subsection{Relativistic tunnel ionization}

Consider now some limiting cases. In the limit of tunnel ionization $\eta \ll 1$ we reproduce the static result of Refs. \cite{PMK97,PMK98} and obtain the first frequency correction
\beq{tunnel}
W^{{\rm tunnel}} &\propto& \exp \left\{- \frac{F_S}{F} \Phi \right\}\, \nn \\
\Phi &=&\frac{2 \sqrt{3} (1-\alpha_0^2)^{3/2}}{\alpha_0} -
\frac{3 \sqrt{3} (1-\alpha_0^2)^{5/2}}{5 \alpha_0} \eta^2 (1-g^2/3)
+ O(\eta^4),
\eeq
where $F_s=m^2c^3/e\hbar=1.32 \cdot 10^{16} {\rm V}/{\rm cm}$ is the Schwinger field of quantum electrodynamics \cite{Schwinger} and $\al_0=(\veps_0+\sqrt{\veps_0^2+8})/4$. In the nonrelativistic regime, $\veps_b=E_b/mc^2 \ll 1$, the parameter $\al_0=1-\veps_b/3+\veps_b^2/27$ and the probability of nonrelativistic tunnel ionization including the first relativistic and frequency corrections reads
\beq{tunnelnon}
W^{{\rm tunnel}} \propto \exp \Bigg\{- \frac{4}{3} \frac{\sqrt{2m}E_b^{3/2}}{e\hbar F} \bigg[ 1 -\frac{\gamma^2}{10}(1-g^2/3) - \frac{E_b}{12 mc^2} \bigg(1-\frac{13}{30} \gamma^2  (1-g^2/3) \bigg) \bigg] \Bigg\}\,.
\eeq
Here the first two terms in the brackets describe the familiar nonrelativistic ionization rate including the first frequency correction, the next two terms are the first relativistic corrections.  It  follows from Eq. (\ref{tunnel}) that the account of relativistic effects increases the ionization rate in comparison with the nonrelativistic rate. However, even for binding energies of the order of the electron rest energy the relativistic correction in the exponent is quite small. 
 In the ``vacuum'' limit Eq. (\ref{tunnelnon}) results into $W \propto \exp \{- {9 F_{\rm S}}/{2 F}\left(1-9/40\eta^2(1-g^2/3)\right)\}$. We find a maximal deviation of about $18\%$ in the argument of the exponential from the nonrelativistic formula.
 Here the ``vacuum'' limit shall not be confused with the pair creation from the vacuum. It is known that there are no nonlinear vacuum phenomena for a plane wave \cite{Schwinger}. In contrast to that we deal here with the ionization of an atom being in rest in the laboratory system of coordinates. Nevertheless, the ``vacuum'' limit should be considered only as the limiting result of the present semiclassical approach where the effects of pair production have been neglected. The polarization of the vacuum becomes important if the binding energy of the atom exceeds the electron rest energy. At the binding energy $E_b=2mc^2$ the single particle picture employed in this paper breaks down ultimately. The electron energy is decreased up to the upper limit for the energy of a free positron, and the threshold energy for the production of an electron-positron pair becomes zero. On the contrary, for a weak relativistic initial state $\veps_b \ll 1$ we expect only a small influence of pair production effects on the ionization process. An appropriate consideration of vacuum polariztion effects can be given only in the framework of quantum electrodynamics. However, this is beyond the scope of the present paper.

\subsection{Relativistic multiphoton ionization}

Consider now the multiphoton limit $\eta \gg 1$. In this case the parameters $\lambda_0=\ln \; ( 2 \gamma/\sqrt{1-g^2})$ (or $\lambda_0=\ln \gamma \sqrt{2 \ln \gamma}$ for $g=\pm 1$) and $\al=1-\veps_b/2 \lambda_0$ and the ionization probability in the relativistic multiphoton limit reads
\beq{multi}
W^{{\rm multi-ph}} &\propto& \exp\left(-\frac{2E_b}{\hbar \omega} f(\gamma \gg 1, g, \veps_b) \right),\\
f(\gamma \gg 1 , g, \veps_b)&=&  \ln \frac{2 \gamma}{\sqrt{1-g^2}} - \frac{1}{2} - \frac{E_b}{8 mc^2 \ln 2 \gamma/\sqrt{1-g^2}} \,,~~~g \neq \pm 1\,,\\
f(\gamma \gg 1, g, \veps_b)&=&  \ln {2 \gamma}{\sqrt{2 \ln \gamma}} - \frac{1}{2} - \frac{E_b}{8 mc^2 \ln 2 \gamma \sqrt{2 \ln \gamma}} \,,~~~~~g = \pm 1\,.
\eeq
Again the first two terms in the function $f(\gamma \gg 1, g,\veps_b)$ reflect the nonrelativistic result \cite{PKM67}, the relativistic effects which lead to an enhancement of the ionization probability are condensed in the third term. 

It has been shown that there is an enhancement of ionization rate in the relativistic theory for both large and small $\eta$.
This should be compared with the results found by Crawford and Reiss.  In their numerical calculations they also found an enhancement of relativistic ionization rate for a circularly polarized field and for $\eta \gg 1$, but for $\eta \ll 1$ their results suggest a strong reduction of the ionization probability \cite{CR94}. For the case of linearly polarized light the ionization rate is found to be reduced by relativistic effects \cite{CR98}. However, Crawford and Reiss studied the above-barrier ionization of hydrogen atom within the strong-field approximation. In contrast to that we have investigated the sub-barrier ionization from a strongly bound electron level, which yields an enhancement of the ionization rate. This enhancement is connected with a smaller initial time $t_0$. As a result the under barrier complex trajectory becomes shorter and the ionization rate increases in comparison with the nonrelativistic theory. Figure 1 shows the relativistic ionization rate Eq. (\ref{relprob})  and the nonrelativistic Keldysh formula as a function of the binding energy $e_b$ and for the case of linear polarization. The figure should be considered only as an illustration of the enhancement effect. The frequency and intensity parameters used for the calculations are still not available for the experimentalists.

\subsection{The case of weak relativistic initial state}

The switch from the multiphoton to the tunnel regime with increasing field strength may be studied in the nonrelativistic limit $\veps_b \ll 1$. Within first order of $\veps_b$ the ionization probability is found to be
\beq{nonrelprob}
W^{{\rm weak-rel}} \propto \exp \left\{- \frac{2E_b}{\hbar \omega} f(\gamma, g, \veps_b \ll 1) \right\},
\eeq
where
$$
f(\gamma,g,\veps_b \ll 1)= f^{(0)}(\gamma,g)+ \varepsilon_b f^{(1)}(\gamma,g)\,.
$$
Here
\be{f0}
f^{(0)}(\gamma, g) = \left(1+\frac{1+g^2}{2 \gamma^2} \right) \lambda^{(0)} -
\left( 1-g^2 + 2g^2 \frac{\th\lambda^{(0)}}{\lambda^{(0)}}\right)
\frac{\sh2\lambda^{(0)}}{4\gamma^2}
\ee
represents the nonrelativistic result \cite{PKM67}, and $\lambda^{(0)}$  satisfies the equation:
$$
\sh^2\lambda^{(0)}-g^2\left(\ch\lambda^{(0)} -
\frac{\sh\lambda^{(0)}}{\lambda^{(0)}}\right)^2= \gamma^2.
$$ 
Besides,
\beq{f1}
f^{(1)}(\gamma,g)= \frac{B}{8\gamma^4}
\left\{
\frac{B+4\gamma^2}{A}
  \left[
\gamma^2 +\frac{1+g^2}{2} -\frac{\ch2\lambda^{(0)}}{2}
     \left(1-g^2 +2g^2 \frac{\th\lambda^{(0)}}{\lambda^{(0)}}\right)
  - g^2 \frac{\th\lambda^{(0)}}{\lambda^{(0)}}
     \left(1- \frac{\sh2\lambda^{(0)}}{2\lambda^{(0)}}\right)
   \right] \right. \nn \\
- \left. \left(1+g^2 + 2g^2\frac{\sh^2\lambda^{(0)}}{\lambda^{(0)2}} +
\frac{1-g^2}{2\lambda^{(0)}} \sh2\lambda^{(0)}
      \right)\lambda^{(0)} 
+ \left(1-g^2 +2g^2 \frac{\th\lambda^{(0)}}{\lambda^{(0)}}\right)
\sh2\lambda^{(0)}
\right\},
\eeq
is the first relativistic correction,
with
\beq{AB}
A &=& \sh2\lambda^{(0)} - 2g^2
\left(\ch\lambda^{(0)} - \frac{\sh\lambda^{(0)}}{\lambda^{(0)}}\right)
\left[\sh\lambda^{(0)} - \frac{1}{\lambda^{(0)}} \left(\ch\lambda^{(0)} -
\frac{\sh\lambda^{(0)}}{\lambda^{(0)}} \right)\right],\\
B &=& 1 + g^2 -2g^2 \frac{\sh^2\lambda^{(0)}}{\lambda^{(0)2}} -
\frac{1-g^2}{2\lambda^{(0)}}\sh2\lambda^{(0)}.
\eeq
 Equation (\ref{nonrelprob}) is valid in the whole $\gamma$-domain, i.e., in the multiphoton regime $\gamma < 1$ as well as in the tunnel limit $\gamma > 1$. For small adiabatic parameters, i.e., $\gamma \to 0$, it coincides with Eq. (\ref{tunnelnon}); in the case of large $\gamma \to \infty$ it transforms to Eq.(\ref{multi}). We mention that Eq. (\ref{nonrelprob}) reproduces the full relativistic formula Eq. (\ref{relprob}) with very high accuracy for $E_b < mc^2$.

 The expression for the  rate of ionization of a weak relativistic initial state essentially simplifies in the case of {\em linear} polarization. Then we have
\beq{nonrelprob-linear}
W^{{\rm weak-rel}} &\propto& \exp \Bigg\{ - \frac{2 E_b}{\hbar \om} f(\gamma, g=0, \veps_b \ll 1) \Bigg\}\,, \nn \\
f(\gamma, g=0, \veps_b \ll 1)&=& \arsinh \gamma + \frac{1}{2 \gamma^2} \left[ \arsinh \gamma - \gamma \sqrt{1+ \gamma^2} \right] - \veps_b \frac{\gamma^4 + \gamma^2 - 2 \gamma \sqrt{1+\gamma^2} \arsinh \gamma + \arsinh^2 \gamma}{8 \gamma^4 \arsinh \gamma} \,.
\eeq
The terms in $f(\gamma, g=0, \veps_b \ll 1)$ which do not vanish as $\veps_b \to 0$ represent the nonrelativistic quasiclassical ionization rate found by Keldysh \cite{Keldysh}; the terms proportional to $\veps_b$ are the first relativistic correction to the Keldysh formula.

\section{Relativistic photoelectron spectrum}

Consider now the modifications of the energy spectrum induced by relativistic effects. First we will characterize the most probable final state of the ejected electron. The classical nonrelativistic barrier-supression ionization predicts a nonzero leaving velocity of the photoelectron. However, relativistic effects as well as frequency corrections modify this result of the classical BSI picture.  In the relativistic semiclassical theory employed in this paper we may set the constants $a_1=0$ and $a_2=(g/\eta)(\sh\lambda_0/\lambda_0)$ according to Eqs. (\ref{a1}) and (\ref{a2}).  From  Eqs. (\ref{trajec}) we obtain then for the most probable emission velocity in the laboratory system of coordinates
\beq{leave}
v_{x,{\rm leaving}} &=& c \frac{1-\alpha^2 +\frac{g^2}{\eta^2}
\left(1- \frac{\sh\lambda_0}{\lambda_0}\right)^2}
{1+\alpha^2 +\frac{g^2}{\eta^2}
\left(1- \frac{\sh\lambda_0}{\lambda_0}\right)^2}\,\\
v_{y,{\rm leaving}} &=& 0\,,\\
v_{z,{\rm leaving}} &=& \frac{2\alpha c}{1+\alpha^2 +\frac{g^2}{\eta^2}
\left(1- \frac{\sh\lambda_0}{\lambda_0}\right)^2}
\frac{g}{\eta} \left(\frac{\sh\lambda_0}{\lambda_0} - 1 \right).
\eeq
where $\al$ has to be taken from  the Eq. (\ref{alpha}).
In the tunnel limit ($\eta <<1$) we obtain:
\Beq
\label{vleave} v_{x,{\rm leaving}} &=&c \frac{1-\alpha_0^2}{1+\alpha_0^2}+O(\eta^2)\,,\\
\label{vz} v_{z,{\rm leaving}}&=&gc\eta \alpha_0 \frac{1-\alpha_0^2}{1+\alpha_0^2}+ O(\eta^3).
\Eeq
Here and below $\alpha_0=(\varepsilon_0+\sqrt{\varepsilon_0^2+8})/4$.
The first term in the $x$-component of the leaving velocity is independent from both the frequency and the intensity of the laser light. It coincides with the static result of Mur {\it et al.} \cite{PMK98}.  The leading term in the $z$-component is proportional to the frequency and inverse proportional to the electric field strength of the laser radiation. From Eqs. (\ref{vleave}) and (\ref{vz}) it also follows that the $x$-component of the leaving velocity vanishes in the nonrelativistic limit, whereas the $z$-component has a nonzero nonrelativistic limit. 
For a nonrelativistic atom, we get:
\beq{vnonr}
v_{x,{\rm leaving}} &=& \frac{v^2}{6c}\left\{1   + O(v^2/c^2,\gamma^2) \right\} \,,\\
v_{z,{\rm leaving}} &=& \frac{v}{6}g\gamma \left\{1  + O(v^2/c^2,\gamma^2) \right\} \,,
\eeq
where $v=\sqrt{2E_b/m}$ is the initial "atomic''
velocity of the electron.
In the "vacuum" limit ($\alpha_0 =1/2$), we have:
\beq{vvacuum}
v_{x,{\rm leaving}} &=& \frac{3}{5}c  + O(\eta^2)\,,\\
v_{z,{\rm leaving}} &=& \frac{3}{10} gc\eta + O(\eta^3).
\eeq

It follows from these equations that a strongly bound electron has a relativistic emission velocity along the direction of the laser beam propagation.
 For a nonrelativistic initial state, $\veps_b \ll 1$, the emission velocity along the beam propagation  is small. Nevertheless, the mean emission velocity seems to be the most sensitive measure of the appearance of relativistic effects in the initial states. In Fig. 2 the $x$-component of the leaving velocity is plotted versus the binding energy of the initial state. Though we have choosen the same parameters of the laser beam as in Fig. 1 it should be mentioned that the dependence of the emission velocity $x$-component on the laser parameters is rather weak. The main parameter determining the leaving velocity along the propagation of the laser beam is the binding energy of the atom.

From Eqs. (\ref{alpha})-(\ref{a2}) we also obtain the most probable value for each component of the final state drift momentum (which is the full kinetic momentum minus the field momentum). We put $a_1=p_{y,m}/mc$, $a_2=p_{z,m}/mc$ and $\al=(-p_{x,m}+\sqrt{m^2 c^2 +p_{x,m}^2+p_{y,m}^2+p_{z,m}^2})/mc$ and get 
\Beq
\label{mom} p_{x,m} &=& \frac{mc}{2\alpha}
\left\{1-\alpha^2 + \frac{g^2}{\eta^2}
\label{py} \frac{\sh^2\lambda_0}{\lambda^2_0} \right\}, \\
p_{y,m} &=& 0\, \\
\label{pz} p_{z,m} &=& mc \frac{g}{\eta}
\frac{sh\lambda_0}{\lambda_0}\,
\Eeq
The leading terms in the tunnel limit ($\eta <<1$) read
\beq{momtun}
p_{x,m} &=& \frac{mc}{2\alpha_0}\frac{g^2}{\eta^2}\,,\\
p_{z,m} &=& mc \frac{g}{\eta}\,.
\eeq
For a nonrelativistic initial state and within the tunnel regime ($\gamma \ll 1$) we obtain
\beq{momnonrel}
p_{x,m}&=&\frac{e^2 F^2 g^2}{2 \om^2 m c}\left(1+\frac{\gamma^2}{3}\frac{g^2+1}{g^2}\right)\,,\\
p_{z,m}&=& \frac{e F}{\om m} g \left(1+\frac{\gamma^2}{6}\right)\,,
\eeq
where we have given the leading terms and the first frequency corrections.

From Eqs. (\ref{mom})-(\ref{pz})  we easily obtain the most probable angle of electron emission. Denote by $\theta$ the angle between the polarization plane and the direction of the photoelectron drift motion; and by $\varphi$  the angle between the projection of the electron drift momentum onto the polarization plane and the smaller axis of the polarization ellipse. In the case of a nonrelativistic atom the most probable angles read
\be{angle2}
\tan \theta_m=\frac{p_{x,m}}{|p_{z,m}|}=\frac{e F|g|}{2mc\om} \left(1+\frac{g^2+2}{g^2}\frac{\gamma^2}{6} \right)\,,~~~\varphi_m=0\,.
\ee
We conclude that relativistic effects produce a nonzero component of the mean electron drift momentum along the axis of beam propagation. As a result the mean angle of electron emission is shifted to the forward direction. However, in the case of linear polarization the appearance of a nonzero $x$-component of the photoelectron drift momentum is connected with relativistic effects in the initial state. The latter are typically small except the case of ionization from inner shells of heavy atoms.
Notice that for the nonrelativistic atom the most probable value for the drift momentum components as well as the expression for the peak of the angular distribution coincide with the corresponding expressions within the BSI model (see Sec. \ref{BSI}) if one neglects the frequency corrections.

Consider now the relativistic final state spectrum, i.e., the momentum distribution near the most probable final state drift momentum.  The calculations will be restricted to the tunnel regime $\gamma \ll 1$. Assuming weak relativistic effects in the initial state, $\veps_b \ll 1$, and putting $\delta p_x=\left(p_x-p_{x,m} \right) \ll mc$, $\delta p_z=\left(p_z-p_{z,m} \right) \ll mc$ and $p_y \ll m c$, one obtains
\beq{pprop}
W_p &\propto& W^{{\rm tunnel}} \exp \Bigg[ -\frac{\gamma}{\hbar \om} \frac{\left[ \delta p_x ^2  -2 \delta p_x \delta p_z \eps g  + \delta p_z^2 \left(1+2\eps^2 g^2 +{\eps^4 g^4}/{4}\right)\right]}{m \; \left(1+\eps^2 g^2/2 \right)^2} \Bigg]\,\nn \\
&& \cdot \exp \Bigg[- \frac{p_{y,m}^2}{3m} \frac{\gamma^3(1-g^2)}{\hbar \om}-  \frac{p_{y,m}^4}{4m^3c^2 \left(1+\eps^2 g^2/2 \right)^2} \frac{\gamma}{\hbar \om} \Bigg]\,,
\eeq
where $W^{{\rm tunnel}}$ is the total ionization rate Eq. (\ref{tunnelnon}) in the weak relativistic tunnel regime. In Eq. (\ref{pprop}) only the leading contributions in $\delta p_x$ and $\delta p_z$ have been given; in the $p_y$ distribution an additional relativistic term proportional to $p_y^4$ has been maintained which becomes the leading term in the case of static fields with $\gamma=0$. In the non-relativistic limit $\eps \ll 1$ and $p \ll c$ we reproduce the results of Ref. \cite{PKM67}. For the cases of linear  ($g=0$) and circular ($g = \pm 1$) polarization our results are in agreement with recent derivations  of Krainov \cite{K98,K99}.

The first exponent in Eq.(\ref{pprop}) describes  the momentum distribution in the plane perpendicular to the major axis of the polarization plane. In the nonrelativistic theory ($\eps=0$) the width of the momentum distribution in $p_x$ coincide with the width of the $p_z$ distribution. The relativistic effects (which are measured by $\eps g$) destroy this symmetry in the (x,z)-plane. The distribution of $p_x$ becomes broader, the $p_z$ distribution becomes narrower. We also mention the appearance of a cross term proportional to the product $\delta p_x \delta p_z$ which is absent in the nonrelativistic theory. In Fig. 3 the distribution of the projection of the photoelectron drift momentum on the axis of the beam propagation is shown. We consider electrons which are produced in the  creation of  ${\rm Ne}^{8+}$ ($E_b=239\, {\rm eV}$) ions by an elliptically polarized laser radiation with wave length $\lambda=1.054 \,\mu {\rm m}$, field strength $2.5 \times 10^{10} \,{\rm V/cm}$ and ellipticity $g=0.707$. The relativistic momentum distribution is compared with the distribution of nonrelativistic theory. From the figure we see that the main effect is the shift of the maximum of the momentum distribution, the broadening remains small for the parameters we have considered.

 The first term in the second exponent of Eq.(\ref{pprop}) determines the nonrelativistic energy spectrum for the low energetic electrons moving along the major polarization axis, 
whereas the second, relativistic term  becomes important for the high energy tail. A detailed analysis of the photoelectron spectrum will be given elsewhere \cite{O99}.

In conclusion,  in this paper  relativistic phenomena for the ionization of an atom in the presence of intense elliptically polarized laser light have been considered.  The cases of  relativistic classical above-barrier and semiclassical sub-barrier ionization have been investigated. Simple analytic expressions for the  ionization rate and the relativistic photoelectron spectrum have been obtained. These expressions apply for relativistic effects in the initial state as well as in the final state. We have shown that relativistic initial state effects lead to a weak enhancement of the ionization rate in the sub-barrier regime. The mean emission velocity has been shown to be a more sensitive measure for the appearance of relativistic effects in the initial state. The more important relativistic final state effects may cause a sharp increase of the electron momentum projection along the propagation of elliptically polarized laser light. This results in a shift of the most probable angle of electron emission to the forward direction.

  Finally, 
the expressions obtained in this paper within exponential accuracy may be improved by taking into account the Coulomb interaction through the perturbation theory. The results of this paper may be also used in nuclear physics and quantum chromodynamics.

\section{Acknowledgements}

This research was partially supported by the Deutsche Forschungsgemeinschaft (Germany).

\newpage

\begin{center}
{\bf FIGURE CAPTIONS}
\end{center}

\begin{description}

\item[(Figure 1)] Absolute value of the logarithm of the ionization rate $-\ln\; W$ versus the binding energy of initial level $e_b=E_b/mc^2$. The solid line shows the relativistic rate Eq.(\ref{relprob}), the dashed line is the nonrelativistic Keldysh formula (Eq. (\ref{nonrelprob}) without the relativistic correction term). The curves are shown for a frequency $\om = 100$ and an intensity $I=8.5 \cdot 10^{7}$ (in a.u.).
\item[(Figure 2)] The $x$-component of the emission velocity {$v_x/c$} versus the binding energy of initial level {$e_b=E_b/mc^2$}.  The emission velocity in the nonrelativistic theory is zero. The curve is shown for a frequency $\om = 100$ and an intensity $I=8.5 \cdot 10^{7}$ (in a.u.).
\item[(Figure 3)]  Spectrum of the electron momentum projection along the beam propagation for electrons produced in the creation of ${\rm Ne}^{8+}$ by an elliptically polarized laser radiation with wave length $\lambda=1.054 \,\mu {\rm m}$, field strength $2.5 \times 10^{10} \,{\rm V/cm}$ and ellipticity $g=0.707$; the relativistic spectrum is taken from Eq. (\ref{pprop}), the nonrelativistic one is Eq.  (\ref{pprop}) with $\eps=0$.

\end{description}

\newpage

\begin {figure} [h] 
\unitlength1mm
  \begin{picture}(155,160)
\put (0,10){\psfig{figure=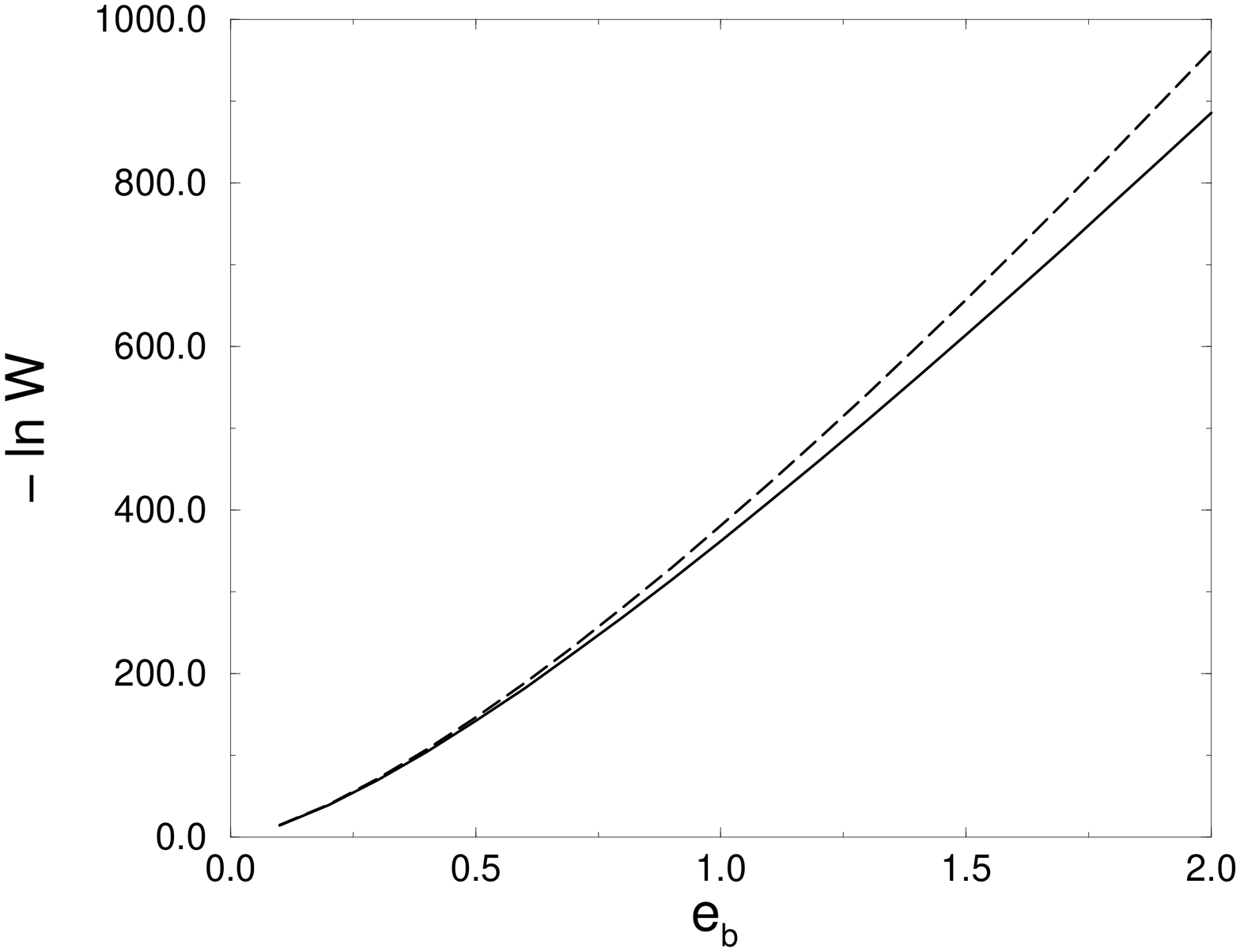,width=15.0cm,height=13.0cm,angle=-90}}
 \end{picture}\par
\caption{}
\end{figure}

\newpage

\begin {figure} [h] 
\unitlength1mm
  \begin{picture}(155,160)
\put (0,10){\psfig{figure=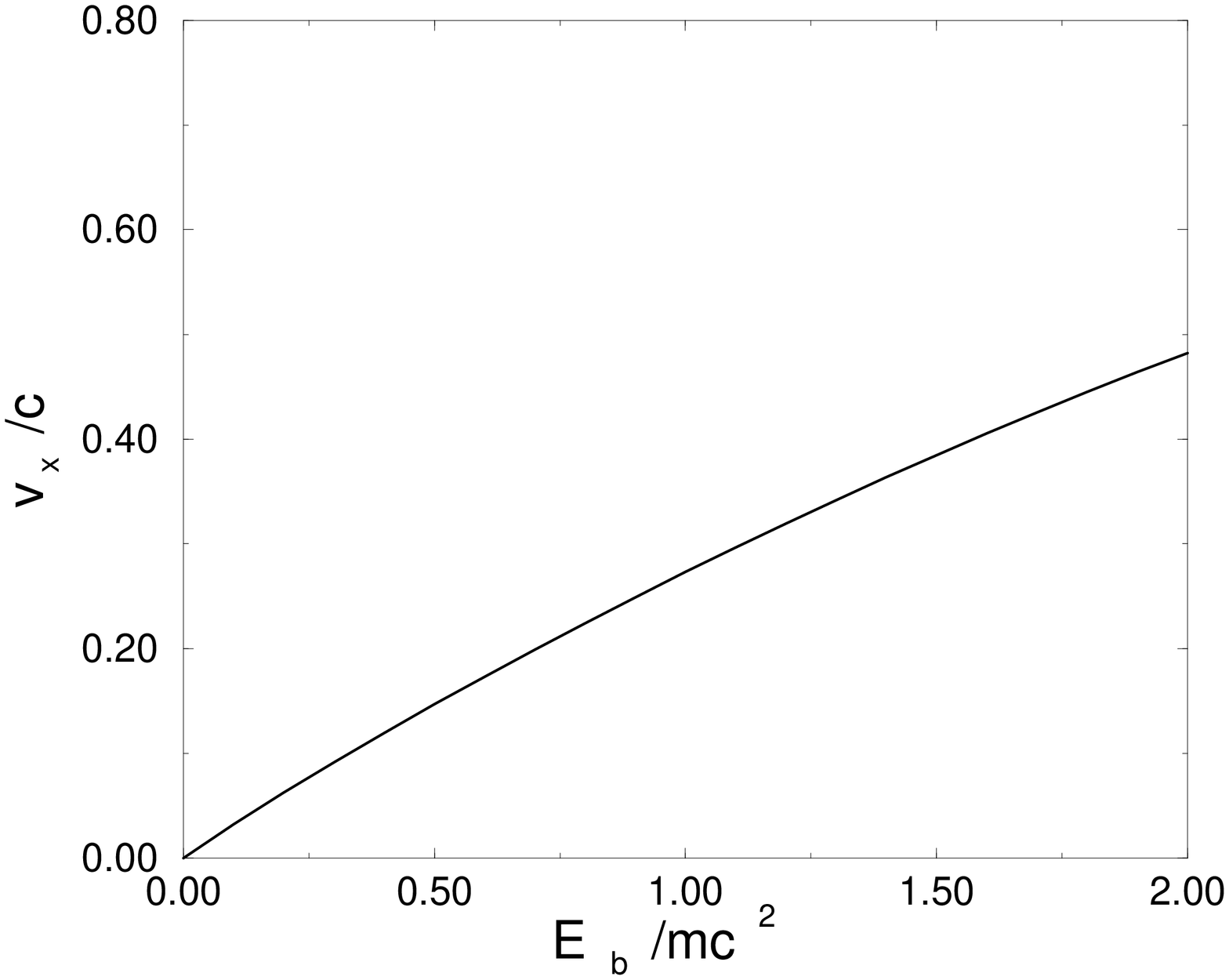,width=15.0cm,height=13.0cm,angle=-90}}
 \end{picture}\par
\caption{}
\end{figure}

\newpage

\begin {figure} [h] 
\unitlength1mm
  \begin{picture}(155,160)
\put (0,10){\psfig{figure=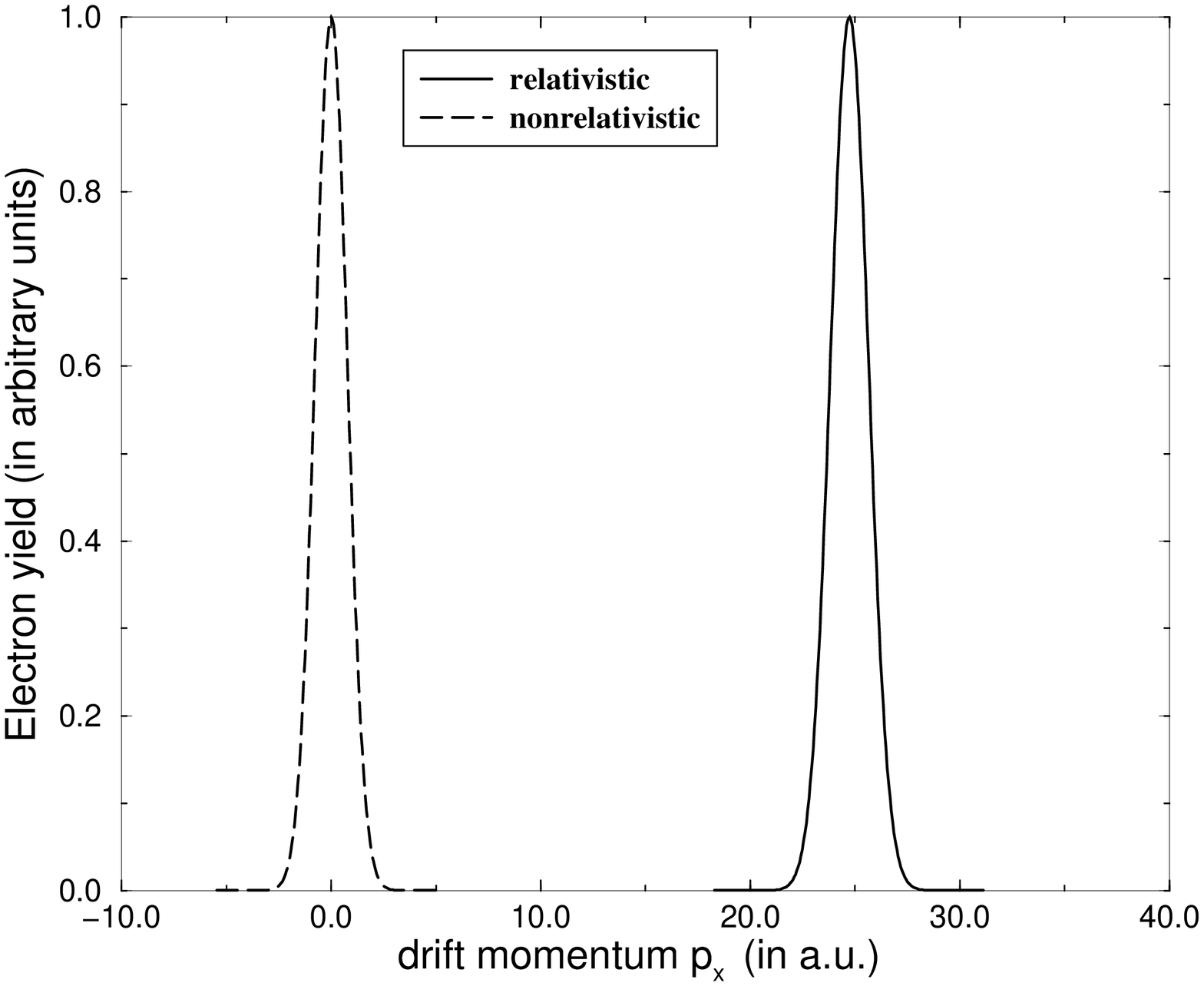,width=15.0cm,height=13.0cm,angle=-90}}
 \end{picture}\par
\caption{}
\end{figure}

\end{document}